\documentclass[conference]{IEEEtran}
\makeatletter
\def\ps@headings{%
\def\@oddhead{\mbox{}\scriptsize\rightmark \hfil \thepage}%
\def\@evenhead{\scriptsize\thepage \hfil \leftmark\mbox{}}%
\def\@oddfoot{}%
\def\@evenfoot{}}
\makeatother
\usepackage[left=0.7in,right=0.7in,bottom=1.04in,top=0.7in]{geometry}
\usepackage{multicol}
\usepackage{nopageno}
\usepackage{color}
\usepackage{soul}
\usepackage{amsmath}
\usepackage{graphics}
\usepackage{setspace}
\usepackage{cite}
\usepackage{latexsym}
\usepackage{float}
\usepackage{epsfig}
\usepackage{multirow}
\usepackage[table,xcdraw]{xcolor}
\usepackage{cite,cases,url}
\usepackage{amssymb}
\usepackage{graphicx}
\usepackage{epstopdf}
\usepackage{balance}
\usepackage{subcaption}
\usepackage{enumerate}
\usepackage{array}
\usepackage{algorithmic}
\usepackage{algorithm}
\usepackage{enumitem}
\usepackage[normalem]{ulem}
\useunder{\uline}{\ul}{}

\newcolumntype{L}{>{\centering\arraybackslash}m{5cm}}
\newcolumntype{K}{>{\centering\arraybackslash}m{6cm}}
\newcolumntype{P}{>{\centering\arraybackslash}m{2.3cm}}
\newcolumntype{M}{>{\raggedright\arraybackslash}m{2cm}}
\newcolumntype{N}{>{\raggedright\arraybackslash}m{2.5cm}}

\begin{document}

\title{
AutoSec-RAN: Specification-Grounded LLM Fine-Tuning for Automated Security Analysis of O-RAN
}
\author{
\IEEEauthorblockN{Anoop Mishra, Joshua Moore, Aly Sabri Abdalla, and Vuk Marojevic \\
} 
\normalsize\IEEEauthorblockA{Dept. of Electrical and Computer Engineering,  Mississippi State University, USA}
\normalsize\IEEEauthorblockA{Emails:  \{am5103; jjm702; asa298;  vuk.marojevic\}@msstate.edu}
}
\maketitle
\begin{abstract}
  This paper presents AutoSec-RAN, a specification-grounded large language model (LLM) framework for automated security triage of Open Radio Access Network (O-RAN) 
  testing. Unlike approaches relying on static rule sets or privileged access to raw packet captures, AutoSec-RAN fine-tunes Mistral-7B-Instruct through two Low-Rank Adaptation (LoRA) stages over a curated corpus of 69 O-RAN, 3GPP, and adjacent standards documents, then selectively augments inference with retrieval-augmented generation (RAG) over the same specification index. At inference, the model receives only an investigator-style question and a scenario metadata block, simulating an automated analyst operating without raw packet evidence. AutoSec-RAN is validated on a purpose-built 5G/O-RAN testbed across labeled fronthaul distributed denial of service 
  replay scenarios. A Gemma-4-31B judge LLM scores each candidate's answer on faithfulness, correctness, and actionability relative to a reference authored from full packet capture 
  and log evidence. AutoSec-RAN achieves a mean judge overall score of 0.713 across all 313 evaluation items (0.755 restricted to the five supervised taxonomy classes), clearing the 0.70 reporting threshold, and a blended score of 0.778 (0.808 supervised-only) when cosine similarity over 
  Beijing Academy of Artificial Intelligence (BAAI) General Embedding (BGE) 
  is incorporated as a secondary semantic signal. The framework performs strongest on mitigation and threat-model reasoning (with a mean overall of 0.849 and 0.848, respectively) and identifies a measurable gap on the held-out attack-direction class (0.515), a controlled result that exposes a specific training data requirement. These results establish specification-grounded LoRA fine-tuning combined with RAG retrieval as a viable foundation for automated security validation of AI-native O-RAN.
\end{abstract}
\IEEEpeerreviewmaketitle

\section{Introduction}
\label{sec:intro}
The emergence of artificial intelligence (AI)-native Radio Access Networks (RANs), built on the open and disaggregated Open Radio Access Network (O-RAN) architecture, introduces new security challenges. The same programmability that enables intelligent RAN optimization can also expose 
open fronthaul 
interface to adversaries that inject spoofed control plane (C-plane) traffic, launch fronthaul denial-of-service (DoS) attacks, or disrupt timing synchronization without triggering conventional alarm thresholds. Validating security in this environment requires correlating heterogeneous evidence drawn from the Enhanced Common Public Radio Interface (eCPRI) payload metadata, wire-direction labels, receive-window counters, and volumetric statistics in ways that static rule sets and signature-based monitors cannot scale to address. As 5G deployments mature and AI-native RAN functions proliferate, the gap between the threat surface and available automated validation tooling grows wider~\cite{abdalla2023end,Aly_ORAN}.

  Large language models (LLMs) offer a path toward closing that gap. Their capacity for analyst-style reasoning over structured technical evidence raises the question of whether a specification-grounded, fine-tuned model can serve as an automated validation engine producing actionable triage judgments without requiring privileged access to RAN internals or real-time packet captures at inference time. The O-RAN fronthaul presents particular challenges for this approach: The governing specifications
  span O-RAN WG11 threat modeling documents~\cite{oran_wg11_threat_modeling}, 3GPP TS 33.501
  cryptographic requirements~\cite{3gpp_ts33501_v18_6}, and eCPRI framing
  conventions~\cite{ecpri_v2}, and an effective model must integrate
 all of these sources to distinguish spoofing from misconfiguration and transient degradation from benign load variation.

\begin{figure}[t]
      \centering
      \includegraphics[width=0.75\linewidth]{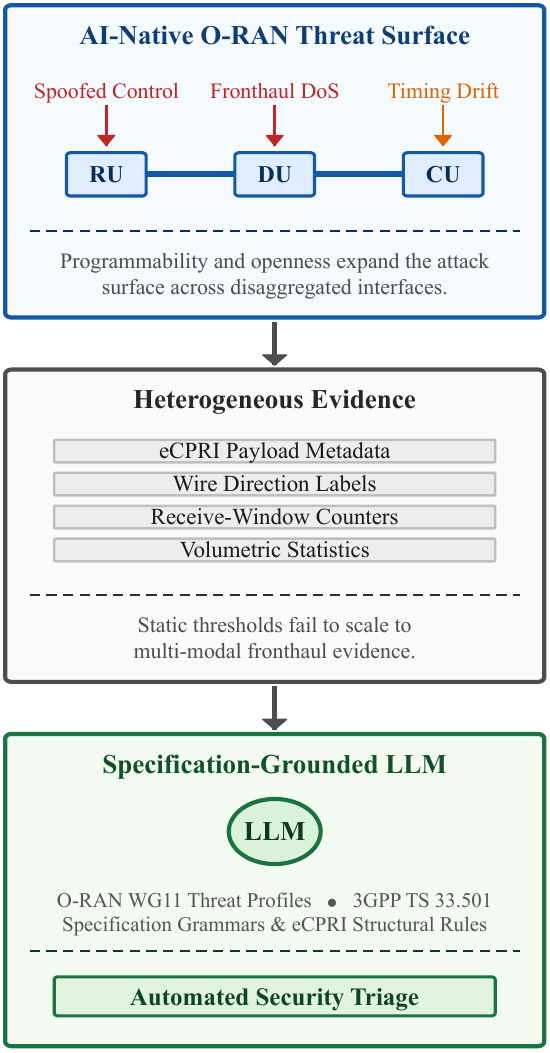}
      \caption{Specification-grounded LLM framework for automated O-RAN security triage. The model receives only a structured question and a seven-field metadata block
  at inference time.}
      \label{fig:arch}
  \end{figure}

  This paper presents AutoSec-RAN (Fig.~\ref{fig:arch}), a two-stage supervised fine-tuning pipeline that addresses this challenge by combining retrieval-augmented generation (RAG) over the curated specification corpus with Low-Rank Adaptation (LoRA) 
  inference on the Mistral-7B-Instruct~\cite{hu2022lora}.
  Stage~1.1 trains on 3,163 analyst question-answer pairs generated from 69 curated O-RAN, 3GPP, and adjacent standards documents\footnote{HuggingFace model ID: \texttt{BAAI/bge-large-en-v1.5}} using a parent-child chunking strategy. Stage~1.2 warm-starts from the Stage~1.1 checkpoint and augments training with
  1,200 observation-driven pairs from the 
  5G-based O-RAN 
  dataset~\cite{zadeh2025descriptor}, extending empirical attack coverage without discarding specification grounding.

  We validate AutoSec-RAN on a purpose-built 5G/O-RAN testbed across labeled Fronthaul Distributed DoS (DDoS) replay scenarios collected using the Soft T-UE platform~\cite{softue}. At evaluation time, the model receives only an investigator-style question and a seven-field scenario metadata block, simulating an automated analyst operating without raw packet evidence. A Gemma-4-31B judge LLM scores each candidate's answer on faithfulness, correctness, and actionability relative to a reference authored from full packet capture (PCAP) and log evidence; a cosine similarity signal over the 
  Beijing Academy of Artificial Intelligence (BAAI) General Embedding (BGE) 
  provides a complementary semantic-overlap check. 
 Our contributions are as follows:

  \begin{itemize}[leftmargin=*,topsep=1pt,itemsep=1pt]
  \item A reproducible synthetic pair generation methodology grounding LLM fine-tuning in O-RAN WG11,
  3GPP, and adjacent standards security specifications.
  \item A multi-axis evaluation rubric separating hallucination avoidance, technical correctness, and analyst utility into independently interpretable signals.
  \item A class-level performance breakdown across six security reasoning taxonomy classes, revealing which aspects of fronthaul threat analysis transfer from specification supervision and which require targeted empirical augmentation.
    \item A specification-driven inference mode that produces
   structured triage output from scenario metadata, without access to PCAP, RAN/radio unit (RU) 
   logs, or other raw packet evidence, positioning the model as a security scenario triage assistant rather than a packet-level classifier or anomaly detector.
  \end{itemize}

\section{Prior Work}
  Prior work at the intersection of LLMs and O-RAN testing organizes around three recent contributions. ORAN-Bench-13K~\cite{oranbench} constructs a 13,000-question
  benchmark to assess the O-RAN domain knowledge of pretrained LLMs, including GPT-4, Gemini, and RAG-augmented variants, establishing that off-the-shelf models leave
  substantial accuracy gaps on specification-grounded questions without task-specific fine-tuning. AI5GTest~\cite{ai5gtest} targets functional conformance through a
  cooperative three-LLM pipeline: A generative model produces expected procedural flows from O-RAN and 3GPP specifications, a validation model cross-references live
  signaling messages against those flows to detect deviations, and a debug model performs root cause analysis on anomalies, together reducing test execution time while
  maintaining high accuracy. ARCANE~\cite{arcane} takes an active security posture by using an LLM to extract finite state machine models from O-RAN specifications and
  then driving a semantics-aware fuzzer over OAI5G, uncovering nine root 
  defects with serious security implications. Subsequent work has extended LLM use into runtime O-RAN security along complementary directions. 6G-XSec~\cite{xsec} pairs an unsupervised deep-learning detector with a pretrained LLM
  that explains flagged anomalies at the cellular edge, and 5G-SPECTOR~\cite{spector} provides rule-based Layer 3 attack detection as an O-RAN-compliant xApp. Most recently, an LLM-based zero-shot pipeline for blind DoS detection over Radio Resource Control (RRC)/Non-Access Stratum (NAS) traffic has been proposed~\cite{blinddos}, relying on short natural-language attack descriptions supplied in-context rather than on specification grounded
  fine tuning. On the data side, TSpec-LLM~\cite{tspecllm} releases an open corpus of 3GPP Releases 8--19 to support retrieval-augmented question answering over telecommunications standards.
  
  AutoSec-RAN differs from all of these works along four axes: (i)~specification-grounded LoRA fine-tuning rather than zero-shot prompting~\cite{blinddos} or
  pretrained-explainer chaining~\cite{xsec}; (ii)~standalone analyst-style triage prose produced from metadata block, without packet or telemetry access at inference, rather than
  detector-coupled explanations~\cite{xsec} or signature-based alerts~\cite{spector}; (iii)~a multi-axis interpretable judge rubric (faithfulness, correctness, actionability) complemented by a BGE cosine semantic check, rather than scalar accuracy or F1 score; and (iv)~end-to-end validation on a physical 5G/O-RAN testbed. 

\section{AutoSec-RAN Testing Methodology and Components}
\label{sec:services}
\begin{figure*}
      \centering
      \includegraphics[width=0.68\linewidth]{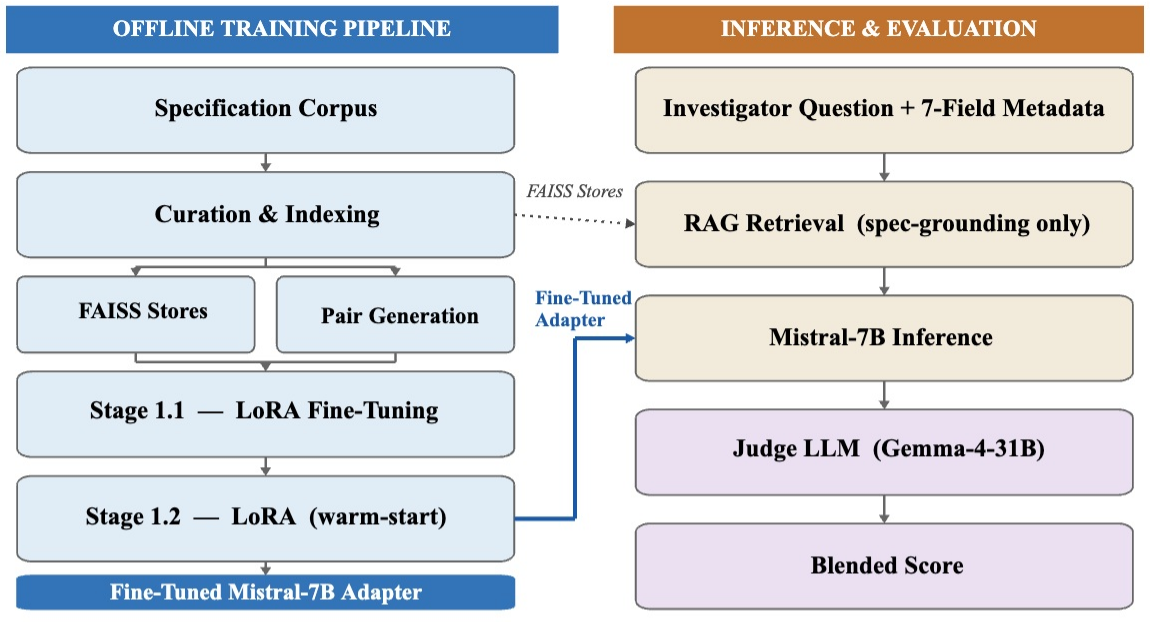}
      \caption{AutoSec-RAN pipeline architecture. The offline training pipeline (left) curates 69 security-relevant O-RAN, 3GPP, and related standards, indexes them into fronthaul and 5G-core FAISS vector stores using parent–child chunking with BGE-large-en-v1.5 embeddings, generates 5,448 analyst-style question-answer pairs with Gemma-4-31B, and fine-tunes Mistral-7B-Instruct via two-stage LoRA. The inference/evaluation pipeline (right) performs RAG-based retrieval from the same FAISS indexes, serves the Stage 1.2 LoRA adapter on a 4-bit NF4-quantized Mistral-7B backbone with bf16 compute, and evaluates outputs using a Gemma-4-31B judge for faithfulness, correctness, and actionability.}
      \label{fig:pipeline}
  \end{figure*}

This section describes the end-to-end pipeline of AutoSec-RAN, covering 
the specification corpus curation and indexing strategy, the synthetic training pair generation process, the two-stage fine-tuning procedure, and the evaluation framework used to score model outputs against analyst-authored references. Fig.~\ref{fig:pipeline} summarizes the offline training pipeline (left) and the inference and evaluation pipeline (right). 

  \subsection{Specification Corpus: Curation and Indexing}
  The retrieval corpus is built from a reviewed keep-list of 69 security-relevant documents retained  from 85 candidate paths spanning O-RAN WG11 threat modeling  specifications~\cite{oran_wg11_threat_modeling}, 3GPP TS 33.501 cryptographic  requirements~\cite{3gpp_ts33501_v18_6}, and O-RAN open-fronthaul security technical reports~\cite{oran_wg11_open_fh_security}, with complementary National Institute of Standards and Technology (NIST), European Telecommunications Standards Institute (ETSI), Internet Engineering Task Force (IETF), and GSM Association (GSMA) security material. Documents excluded from the keep-list were judged insufficiently security-specific or redundant with retained sources. Retained documents are extracted and split using a parent-child retrieval strategy with BGE embeddings. Parent chunks are sized at 1,800 characters with 200-character overlap and serve as the context window supplied to the LLM during RAG inference. Child chunks are sized at 350 characters with 50-character overlap and are indexed in a Facebook AI Similarity Search (FAISS) vector store for similarity search at query time. The current corpus contains 2,832 fronthaul parent chunks and 1,343 5G-core parent chunks, for 4,175 total parent chunks across the two domain stores. 

  \subsection{Training Data Generation}
  The specification corpus composed of training pairs is generated by Gemma-4-31B-IT\footnote{HuggingFace model ID: \texttt{google/gemma-4-31b-it}} from parent chunks of at least 400 characters. For each qualifying chunk, Gemma produces one investigator-framed question and one analyst-style answer under one of five rotating taxonomy classes:
  \textit{threat\_model}, \textit{detection\_signals}, \textit{mitigation}, \textit{spec\_grounding}, and \textit{scenario\_severity\_expectation}. The
  \textit{attack\_direction\_reasoning} class is intentionally excluded from specification training to create a held-out evaluation condition that tests generalization
  beyond the supervised distribution. Generated answers must be written in analyst prose, cite short specification identifiers, and avoid literal severity or rank tokens. A consistency audit reviews one in five generated pairs by sampling the source chunk and verifying that the answer is supported by the chunk content rather
  than model priors.

  Stage 1.2 adds observation-driven supervision from the 5G-based ORAN higher-layer flow dataset~\cite{zadeh2025descriptor}. Flow records are stratified across classes, including Synchronize Flag (SYN), Transmission Control Protocol (TCP), PortScan, User Datagram Protocol (UDP), Internet Control Message Protocol (ICMP), and benign traffic, and each row is formatted into a structured observation block containing protocol, service, duration, byte counts, packet counts, connection state, and history fields. Gemma converts these observations into the same analyst question-and-answer format used for specification pairs. Because
  NetsLab labels are treated as indicators rather than authoritative ground truth. All NetsLab pairs are flagged as consistency-only and framed as evidence-based analyst reasoning rather than definitive attack declarations. Each pair carries an instruction, the analyst question, and the analyst answer, together with metadata for taxonomy class, source, cited specification identifiers, and audit flags, and is rendered into Mistral's \texttt{<|system|>}, \texttt{<|user|>}, \texttt{<|assistant|>} chat template at training time. The current manifest contains 5,448 valid pairs in total. Stage 1.1 uses 3,163 specification-analyst pairs. Stage 1.2 uses 4,363 pairs, composed of the same 3,163 specification
  pairs plus 1,200 NetsLab pairs. Validation holds 815 pairs, and test holds 270 pairs under a stratified split by source and taxonomy class.

  \subsection{Two-Stage LoRA Fine-Tuning}

  Both stages fine-tune Mistral-7B-Instruct-v0.3\footnote{HuggingFace model ID: \texttt{mistralai/Mistral-7B-Instruct-v0.3}} using LoRA~\cite{hu2022lora} applied to the query, key, value,
  and output projection matrices of every attention layer. The LoRA configuration uses rank $r = 16$, $\alpha = 32$, and dropout 0.05. Training uses a batch size of 4 with gradient accumulation over 4 steps for an effective batch size of 16, \texttt{bf16} mixed precision, a cosine learning-rate schedule, and paged AdamW with 8-bit quantization. Early stopping is applied on validation loss to prevent overfitting to the synthetic training distribution. Stage 1.1 trains for three epochs at a learning rate of $2\times10^{-4}$, producing a specification-grounded base adapter. Stage 1.2 warm-starts from the Stage 1.1 checkpoint and trains for three additional epochs at a reduced learning rate of $5\times10^{-5}$, preserving the specification grounding established in Stage 1.1 while incorporating the empirical NetsLab signal. The warm-start strategy is preferred over joint training on the combined dataset because it allows Stage 1.1 to establish stable attention patterns over specification language before the more heterogeneous flow-observation pairs are introduced.



\section{Evaluation Framework} 
\label{sec:systen}
This section elaborates on the evaluation framework, rationale behind each evaluation criterion, the process by which reference answers and rubrics are constructed, the implementation of the judge LLM pipeline, and the experimental configuration used to produce the results reported in
  Section~\ref{sec:testbed}.

  \subsection{Evaluation Criteria}

  Security analyst triage quality cannot be captured by a single scalar. A response may be technically accurate yet contain hallucinated citations, or it may reproduce specification language without offering any actionable guidance. AutoSec-RAN, therefore, decomposes the evaluation into four interpretable sub-scores, each targeting a distinct dimension of analyst output quality.

  \begin{itemize}
      \item{Faithfulness} penalizes unsupported or hallucinated claims. In a security context, hallucination is particularly costly because an analyst acting on a fabricated protocol behavior or a nonexistent specification requirement may misconfigure defenses or dismiss a genuine threat.

  \item{Correctness} measures whether the technical claims in the candidate answer agree with the reference answer authored from ground-truth PCAP and log evidence. Because factual accuracy is the primary requirement for analyst-grade output. 

  \item{Actionability} measures whether a security analyst could directly apply the candidate's answer to triage, detection, or mitigation work without additional research. A response that correctly identifies a spoofing condition but provides no guidance on how to detect or contain it has
   limited operational value.

  \item{Coverage} measures whether the candidate addresses the required must-mention concepts specified in the per-item rubric. Coverage is computed for each item, but is excluded from the weighted overall score because rubric must-mention lists vary in length and specificity across taxonomy classes, making cross-item aggregation misleading. Coverage is retained as a diagnostic signal for identifying systematic gaps in conceptual completeness.
  \end{itemize}

  
  \subsection{Evaluation Score}

  Each evaluation unit consists of the model input question, the seven-field scenario metadata block, a reference answer authored from full PCAP and log evidence, and a
  per-item rubric specifying must-mention concepts and must-not-mention terms. The
  judge LLM 
  receives all four components and outputs four sub-scores on the continuous interval $[0, 1]$: \textit{faithfulness} (F),
  measuring whether the candidate avoids unsupported or hallucinated claims; \textit{correctness} (Cr), measuring agreement of technical claims with the reference
  answer; \textit{actionability} (A), measuring whether a security analyst could directly use the answer for triage, detection, or mitigation work; and \textit{coverage}
   (Cv), measuring whether required must-mention concepts are addressed. 

Let $N$ denote the number of evaluation items and, for item $i \in \{1,\dots,N\}$, let $F_i, Cr_i, A_i, Cv_i \in [0,1]$ denote the four judge sub-scores and $c_i \in [-1,1]$ the raw BGE cosine. The per-item weighted overall score is

\begin{equation}
      O_i \;=\; w_F\, F_i \;+\; w_{Cr}\, Cr_i \;+\; w_A\, A_i,
      \label{eq:overall}
  \end{equation}

  \noindent with weights $w_F = 0.3$, $w_{Cr} = 0.5$, $w_A = 0.2$ held fixed across items. Coverage ($Cv_i$) is reported diagnostically but excluded from $O_i$. The set-level mean overall score and the mean raw cosine are

\begin{equation}
      \bar{O} \;=\; \frac{1}{N}\sum_{i=1}^{N} O_i,
      \qquad
      \bar{c} \;=\; \frac{1}{N}\sum_{i=1}^{N} c_i.
      \label{eq:means}
  \end{equation}

  \noindent Because raw cosine spans $[-1,1]$, it is rescaled to the unit interval inside the blended score:

 \begin{equation}
      B \;=\; w_B\, \bar{O} \;+\; (1 - w_B) \cdot \frac{\bar{c} + 1}{2},
      \qquad w_B = 0.7.
      \label{eq:blended}
  \end{equation}

  \noindent An item passes when its overall score meets a fixed threshold $\tau$, which is calibrated to correspond to the minimum quality level at which a trained security analyst would find the response useful for operational decision-making. The set-level pass rate is

  \begin{equation}
      p_\tau \;=\; \frac{|\{\, i \,:\, O_i \geq \tau \,\}|}{N},
      \qquad \tau = 0.70.
      \label{eq:passrate}
  \end{equation}

  \subsection{Reference Answer and Rubric Construction}

  Reference answers are authored offline from the full evidence available for each replay scenario, including the PCAP capture, raw gNodeB and RU logs, outcome labels, and window-level timing statistics. This evidence base is richer than the seven-field metadata block available to the model at inference time, and this gap is intentional. The reference establishes the ground truth that a fully informed analyst would produce, while the model is evaluated on how closely it approaches that standard from the constrained input it receives.

   For each replay scenario, the judge model uses a rubric that specifies two components. The must-mention list identifies short concept strings, such as specification section identifiers or attack class labels, that a complete answer is expected to address. The must-not-mention list identifies specific numerical values or log tokens that are present in the raw evidence but whose inclusion in a candidate answer would indicate data leakage rather than reasoning. 

  \subsection{Judge LLM}

  The judge LLM is Gemma-4-31B-IT, selected for its strong instruction-following behavior on structured scoring tasks and its separation from the candidate model family. At inference time, the judge receives a prompt containing four components in fixed order: the investigator question, the reference answer, the candidate answer, and the per-item rubric. The judge is instructed to output exactly four sub-scores as floating-point values in $[0,
  1]$ followed by a short rationale paragraph. No chain-of-thought scratchpad is exposed in the final output to prevent the rationale from influencing sub-score calibration.

  To reduce positional bias, the order in which the reference and candidate answers appear in the judge prompt is held constant across all items. Judge outputs are parsed with a strict regex that rejects malformed responses. In the event of a parse failure, the item is excluded from the aggregate rather than assigned a default score. No manual corrections are applied to the reported results.
  
  Single-judge evaluation is known to exhibit verbosity, position, and self-preference biases~\cite{zheng2023judging}. AutoSec-RAN adopts three implementation choices intended to mitigate these failure modes: (i)~a fixed component order in the judge prompt (question, reference, candidate, must-mention, must-not-mention) held constant across all items; (ii)~strict regex-based JSON parsing with up to three retries at low temperature, after which malformed responses are flagged as parse failures and excluded from the aggregate rather than assigned a default; and (iii)~no manual correction of judge outputs at any stage. The BGE cosine signal introduced in the next subsection serves as an independent semantic check on the judge's verdicts.

  The complete set of candidate answers is additionally scored by two reference judges from a separate model family, Qwen2.5-32B-Instruct and Qwen2.5-72B-Instruct.\footnote{HuggingFace model IDs:
  \texttt{Qwen/Qwen2.5-32B-Instruct} and \texttt{Qwen/Qwen2.5-72B-Instruct}}
  The candidate answers, reference answers, prompt structure, and the sub-score weights of~\eqref{eq:overall} are held fixed, so that the judge weights are the only variable. Agreement between judges is measured with Spearman's rank correlation over per-item overall scores, which confirms whether the judges order the evaluation items consistently.

  \subsection{Cosine Similarity and Blended Scoring}

  The cosine similarity signal is computed by encoding both the reference answer and the candidate answer with BGE 
  using CLS-token pooling over the final hidden states. Raw cosine similarity spans $[-1, 1]$ and is normalized to $[0, 1]$ as defined in~\eqref{eq:blended} while blending. The blended score defined in~\eqref{eq:blended} combines the mean judge's overall score and the normalized cosine at weights of 0.7 and 0.3, respectively. The cosine similarity score serves as a semantic sanity check. If a candidate's answer performs well according to the judge's criteria but exhibits low cosine similarity, this may suggest that the judge is rewarding plausible-sounding but thematically
  divergent responses. Conversely, a high cosine with low judgment may indicate verbatim repetition of specification language without analytical synthesis. Reporting both signals together surfaces these failure modes more reliably than either signal alone.

\section{Experimental Deployment and Results}
\label{sec:testbed}
This section reports the deployment configuration used during the evaluation campaign, describes the attack scenario execution and replay procedure, presents the
  aggregate and class-level results, and provides qualitative case studies that illustrate representative success and failure modes of the AutoSec-RAN framework.



  \subsection{Testbed Architecture and Data Collection}
 AutoSec-RAN is validated on a purpose-built 5G/O-RAN testbed~\cite{upadhyaya2022prototyping}. The testbed instantiates a disaggregated RAN stack conforming to the O-RAN Alliance reference architecture, with a software-defined radio (SDR)-based gNodeB split across a central unit (CU), a distributed unit (DU), and an RU connected over an eCPRI-based open fronthaul~\cite{ecpri_v2}.
  Labeled PCAP traces are collected under fronthaul DDoS replay scenarios directed at the eCPRI interface using the Soft T-UE framework~\cite{11309908}, producing the 313-item reference set used for evaluation in this paper. Each capture session is paired with raw gNodeB and RU logs, outcome labels indicating whether the RU remained operational or experienced degradation, and window-level statistics including packet arrival timing and receive-window late-packet counts. This richer evidence base is used exclusively to author reference answers and rubrics at evaluation time. 
The analyst has access to high-level scenario descriptors but not to raw packet evidence. That is, no PCAP data or log content is passed to the model at inference time.
  
  \subsection{Deployment Configuration}

  The Stage 1.2 LoRA adapter is attached via PEFT to Mistral-7B-Instruct-v0.3 loaded in 4-bit NF4 quantization with bfloat16 
  compute data type for evaluation inference. The
  RAG pipeline runs as a preprocessing step before model calls in the spec-grounding class only; items in the remaining four supervised classes skip retrieval entirely.
  When retrieval is invoked, the investigator's question is embedded with BGE-large-en-v1.5, the resulting vector is used to query the FAISS child-chunk index, and the top 20
  candidates are re-ranked with a BGE-reranker-large cross-encoder, and the top retrieved child chunks are resolved to their parent chunks, which are prepended to the
  model input ahead of the seven-field scenario metadata block. The judge pipeline runs as a postprocessing step: after the candidate answer is generated, it is
  submitted to Gemma-4-31B-IT alongside the reference answer and the per-item rubric to produce the four sub-scores and a short rationale. Both inference calls are
  stateless; no conversation history or cross-item context is carried between evaluation units. The embedding model used for cosine similarity scoring is the same
  BGE-large-en-v1.5 model used for RAG retrieval, with CLS-token pooling applied over the final hidden states to produce fixed-length sentence embeddings.

  \subsection{Attack Scenario Configuration}

    Each attack scenario on the testbed is executed as a controlled open fronthaul traffic-injection replay targeting the eCPRI interface between the DU and the RU. 
    The attack execution is done under a fixed configuration defined by the seven metadata fields: injection point, attack direction, packet size, data rate, source volume, plane, and link direction. The injection point field specifies which node introduces the malicious traffic, either the DU Network Interface Card (NIC) or the RU NIC side of the open fronthaul link. The attack direction field encodes whether the injected frames travel from the DU toward the RU or in the reverse direction. Packet size, rate in Mbps, and source volume together parameterize the traffic load profile. The plane and link direction fields specify
  whether the target traffic belongs to the C-plane or the user plane, and whether it flows downlink or uplink. Across all scenarios, PCAP captures are recorded at the fronthaul interface and retained alongside gNodeB and RU logs for reference answer authoring~\cite{101145}. 

  \subsection{Aggregate Results}

\begin{figure}[t]
      \centering
      \includegraphics[width=0.95\linewidth]{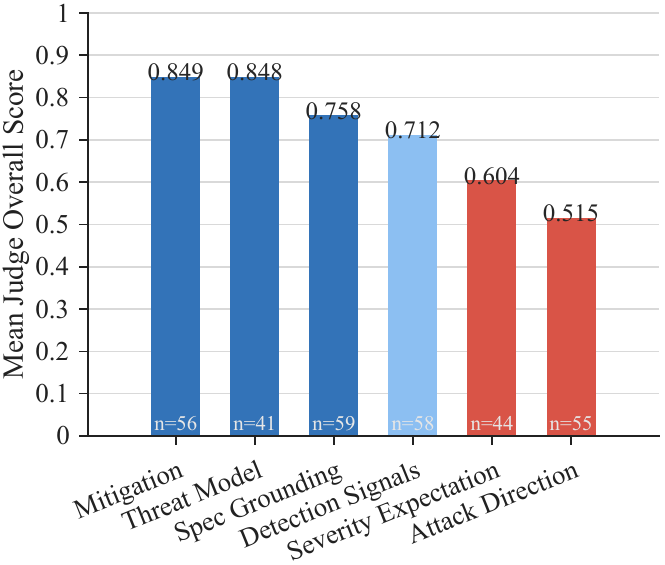}
      \caption{Mean judge overall score per taxonomy class. The sample size per class is annotated within each bar.}
      \label{fig:class_mean_overall}
  \end{figure}

\begin{figure}[t]
      \centering
      \includegraphics[width=0.95\linewidth]{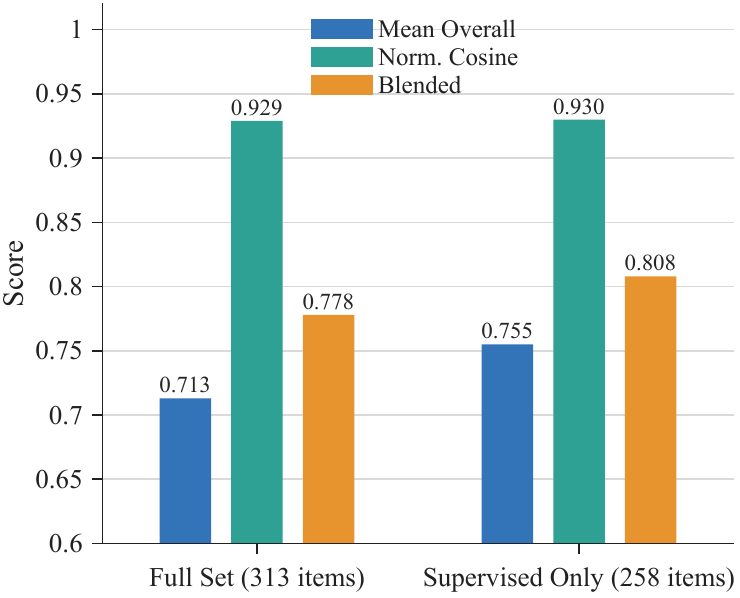}
      \caption{Aggregate score components for the full evaluation set (313 items) and the supervised-class-only subset (258 items, excluding the held-out Attack Direction Reasoning class). Each group shows the mean overall, normalised cosine, and blended score.}
      \label{fig:aggregate_scores} 
  \end{figure}

  Table~\ref{tab:class_results} reports the per-class breakdown, and Fig.~\ref{fig:class_mean_overall} visualizes the mean judge overall score for each taxonomy class. At the aggregate level, AutoSec-RAN achieves a mean judge overall score of 0.713 across all 313 evaluation items, clearing the 0.70 pass
   threshold. The mean raw cosine similarity between reference and candidate answer embeddings is 0.859, which normalizes to 0.929. The full-set blended score, computed from these two signals at weights 0.7 and 0.3 as defined in~\eqref{eq:blended}, is 0.778. Restricting the aggregate to the five taxonomy classes covered by specification training, that is, excluding the intentionally withheld attack-direction reasoning class, the mean overall rises to 0.755, and the blended score rises to 0.808. Fig.~\ref{fig:aggregate_scores} contrasts the two settings on all three score components. The difference between the full-set and supervised-only blended scores,  0.030, provides a direct estimate of the performance cost attributable to the held-out class. 

   Re-scoring the candidate answers with the two reference judges yields mean overall scores of 0.728 and 0.776 and pass rates of 0.760 and 0.744, both above the primary judge. All three judges rank the six taxonomy classes in the same order by mean overall score, and their per-item score orderings are strongly correlated, with Spearman's rank correlation of 0.773 and 0.801 between the primary judge and each reference judge and 0.780 between the two reference judges.
    
    \begin{table}[t]
    \centering
    \vspace{+5 pt}
    \caption{Per-Class Evaluation Results}  
    \label{tab:class_results}
    \renewcommand{\arraystretch}{1.15}

    \resizebox{\columnwidth}{!}{%
     \begin{tabular}{lc cc cc cc}
    \hline
    \textbf{Taxonomy Class} & \textbf{$n$}
      & \multicolumn{2}{c}{\textbf{Gemma-4-31B}}
      & \multicolumn{2}{c}{\textbf{Qwen2.5-32B}}
      & \multicolumn{2}{c}{\textbf{Qwen2.5-72B}} \\
      & & \textbf{Mean} & \textbf{Pass @ 0.70}
      & \textbf{Mean} & \textbf{Pass @ 0.70}
      & \textbf{Mean} & \textbf{Pass @ 0.70} \\
    \hline
    Mitigation                    & 56 & 0.849 & 0.875 & 0.844 & 1.000 & 0.868 & 0.982 \\
    Threat Model                  & 41 & 0.848 & 0.756 & 0.812 & 0.854 & 0.819 & 0.854 \\
    Spec Grounding                & 59 & 0.758 & 0.525 & 0.782 & 0.881 & 0.804 & 0.847 \\
    Detection Signals             & 58 & 0.712 & 0.534 & 0.754 & 0.759 & 0.791 & 0.862 \\
    Scenario Severity Expectation & 44 & 0.604 & 0.341 & 0.712 & 0.659 & 0.712 & 0.500 \\
    Attack Direction Reasoning    & 55 & 0.515 & 0.200 & 0.473 & 0.400 & 0.655 & 0.382 \\
    \hline
    \textbf{Full Set} & 313 & \textbf{0.713} & \textbf{0.537} & \textbf{0.728} & \textbf{0.760} & \textbf{0.776} & \textbf{0.744} \\
    \hline
    \end{tabular}%

    }
    \end{table}

\begin{figure}[t]
      \centering
      \includegraphics[width=0.85\linewidth]{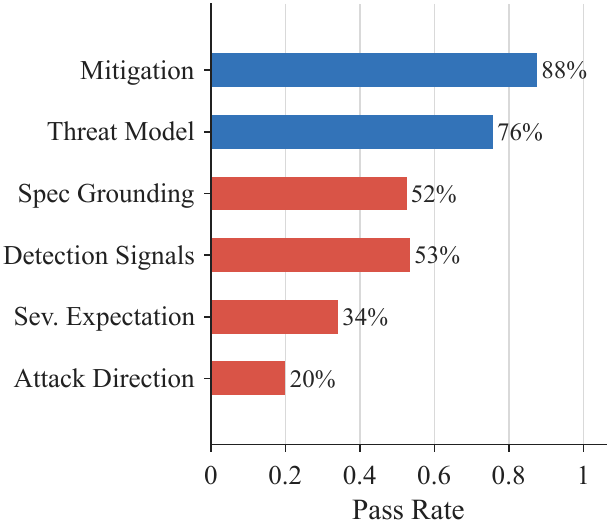}
      \caption{Per-class pass rate at the judge. Classes are sorted in ascending order of pass rate. Mitigation and Threat Model are the only classes that exceed 70\%;
  Attack Direction Reasoning, excluded from specification training by design, achieves the lowest pass rate at 20.0\%.}
      \label{fig:class_pass_rate}
  \end{figure}

  Per-class pass rates vary substantially, as illustrated in~Fig.~\ref{fig:class_pass_rate}. Mitigation and threat model pass on more than three-quarters of items at 0.875 and
  0.756, respectively. Spec grounding and detection signals pass on roughly half at 0.525 and 0.534. Scenario severity expectation passes on about a third at 0.341, and
  attack direction reasoning on a fifth at 0.200. The monotonic relationship between class pass rate and proximity to the specification training distribution is
  consistent across all six classes, confirming that the fine-tuning signal transfers to held-in classes and degrades gracefully rather than catastrophically on the
  held-out class.

  \subsection{Class-Level Analysis}
  \textit{Mitigation and Threat Model:} These two classes benefit most directly from the specification training corpus. O-RAN WG11~\cite{oran_wg11_threat_modeling,oran_wg11_security_protection,oran_wg11_zta} and 3GPP TS 33.501~\cite{3gpp_ts33501_v18_6} contain extensive normative guidance on recommended mitigations including ingress filtering, MACsec, IPsec, and zero-trust interface policies, and the model has been trained to synthesize this guidance into analyst prose. The high faithfulness sub-scores in both classes (mean faithfulness above 0.95 across both) indicate that the model rarely introduces unsupported claims when operating within the scope of the
  specification corpus.

  \textit{Spec Grounding and Detection Signals:} Performance in these two classes is adequate but less consistent. Spec grounding requires the model to cite specific normative requirements by identifier and to distinguish between shall and should obligations, a precision task that the parent-child retrieval strategy supports but does not guarantee when the relevant chunk is not ranked in the top-$k$ retrieved set. Detection signals question the model to reason about the shape of deviation from baseline on specific monitoring axes, and the model frequently
  provides correct axis identification while omitting the expected deviation direction or magnitude. Both failure modes are reflected in correctness sub-scores that average 0.63 for spec grounding and 0.58 for detection signals.

  \textit{Scenario Severity Expectation:} This class requires the model to predict whether a given attack configuration will cause the RU to crash, degrade gracefully, or recover without intervention. The prediction depends on operational experience with attack parameter thresholds that are not present in the specification corpus and only partially represented in the NetsLab flow data. The result is a mean overall of 0.604, driven primarily by low correctness scores on scenarios where the ground-truth outcome is recovery without crash, a nuanced label that the model tends to conflate with more severe outcomes.

  \textit{Attack Direction Reasoning:} This class is excluded from all training data by design. The 0.515 mean overall and 0.200 pass rate represent near-chance performance on questions that require the model to reason about the relationship between the Ethernet wire direction and the
  eCPRI payload direction label, and to classify that relationship as spoofing, legitimate, or misconfiguration. The result validates the experimental design: the model cannot recover this reasoning from specification language alone, establishing a clear and targeted training data
  requirement for the next pipeline iteration.

\begin{figure}[t]
      \centering
      \includegraphics[width=0.95\linewidth]{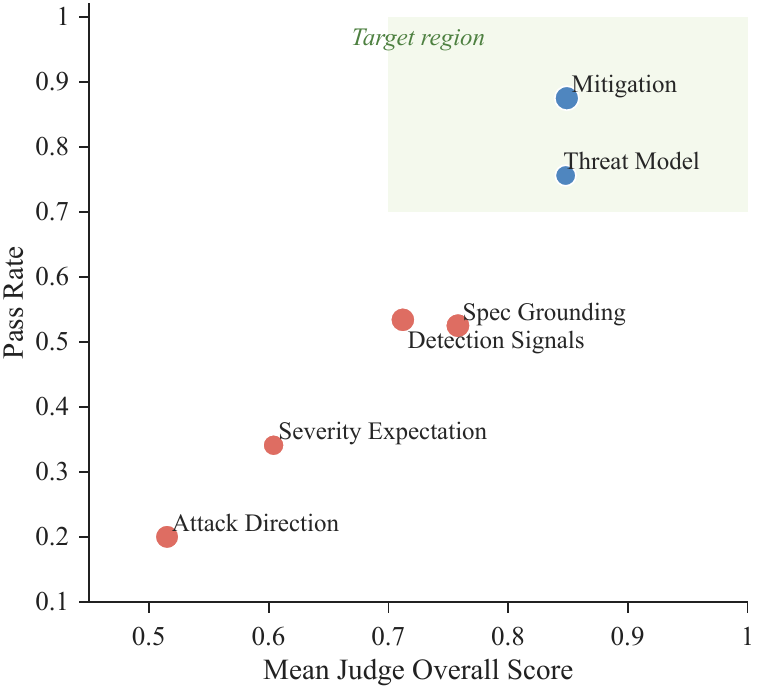}
      \caption{Joint distribution of mean overall score and pass rate across taxonomy classes. Bubble area is proportional to the number of evaluation items in each
  class; the shaded region identifies the target quadrant where both score quality and consistency are acceptable.}
      \label{fig:overall_vs_passrate}
  \end{figure}

Fig.~\ref{fig:overall_vs_passrate} summarizes the joint distribution of mean overall score and pass rate across the six classes and identifies Mitigation and Threat Model as the only classes inside the target quadrant where both score quality and consistency are acceptable.

  \subsection{Qualitative Case Studies}

\begin{figure}[t]
      \centering
      \includegraphics[width=0.95\linewidth]{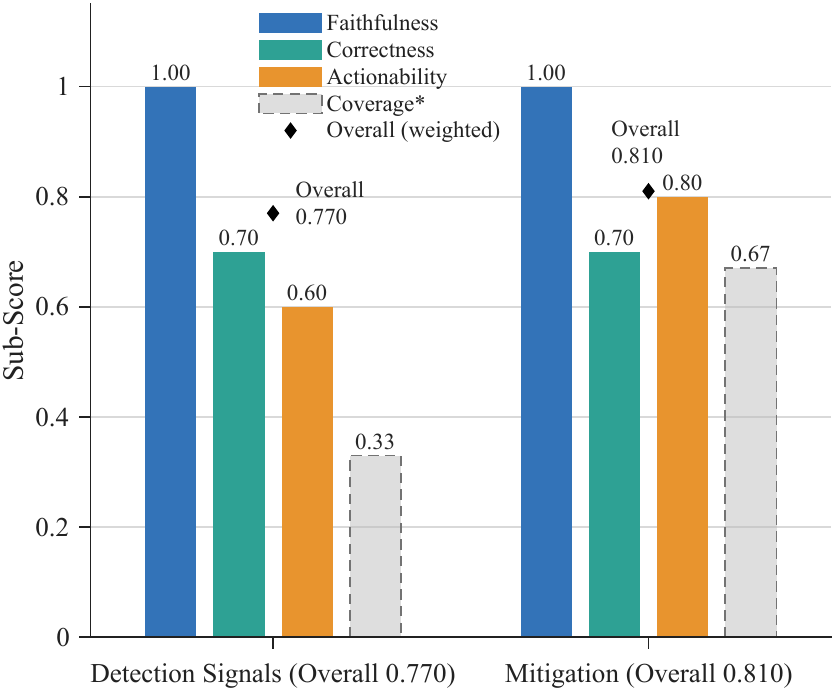}
 \caption{Judge sub-score breakdown for two case study evaluation items from the C-plane downlink RU–DU traffic-injection scenario. Faithfulness, Correctness, and Actionability contribute to the weighted Overall score, while Coverage, shown in hatched gray, is reported only as a diagnostic metric. Diamond markers indicate the final weighted Overall score for each item.}
      \label{fig:case_study_subscores}
  \end{figure}

  
  Two evaluation items from a C-plane downlink RU–DU traffic-injection scenario illustrate the representative performance profile of AutoSec-RAN across different taxonomy classes under the same underlying attack configuration. In this scenario, the baseline PCAP contains long-payload C-plane downlink traffic with a 10 MB source volume, while the generated attack PCAP introduces RU–DU directed traffic replayed at 10 Mbps and injected at the DU-side interface. Fig.~\ref{fig:case_study_subscores} shows the judge sub-score breakdown for both items.

  \subsubsection{Detection Signals Item}

  The question asks which fronthaul detection axes would surface the attack and what shape of deviation versus baseline would be expected on each. The model correctly identifies both the eCPRI payload-versus-wire direction mismatch and the RU receive-window late-packet counters as relevant
   axes, and it references O-RAN WG11 threat modeling in its response, earning a faithfulness score of 1.000 and contributing to a mean overall of 0.770. However, the model characterizes the attack as a man-in-the-middle condition rather than as a spoofing attack, which is the
  classification the reference answer establishes from PCAP evidence. The model also describes what the receive-window counters measure generically rather than predicting the specific deviation shape the question requests, namely an increase in late packets during the attack period
  followed by recovery. These omissions reduce the correctness sub-score to 0.700 and the actionability sub-score to 0.600. The coverage score of 0.330 reflects the failure to use the term spoofing and to treat the eCPRI payload itself as the source of the direction label. The cosine
  similarity of 0.832 confirms that the response is semantically adjacent to the reference without being conceptually equivalent.

  \subsubsection{Mitigation Item}

  The question asks what architectural mitigations should be implemented given the mismatch between wire direction and payload direction. The model correctly recommends IPsec, references O-RAN WG11 and 3GPP TS 33.501, and frames the response around preventing unauthorized frame injection,
  earning a faithfulness score of 1.000 and an actionability score of 0.800, for a mean overall of 0.810. The primary correctness deduction stems from two issues: First, the model recommends mutual TLS for the RU-DU interface, whereas the fronthaul plane at this layer relies on IPsec or
  MACsec rather than TLS, making the suggestion technically imprecise for the eCPRI layer. Second, the model does not mention eCPRI data direction field validation as a required architectural check, which the reference answer identifies as the most direct mitigation for a
  wire-versus-payload mismatch. The coverage score of 0.670 reflects partial satisfaction of the must-mention list, with O-RAN WG11 and 3GPP TS 33.501 cited but eCPRI data direction omitted.

 \subsection{Discussion}

The results establish three properties of AutoSec-RAN: First, specification-grounded LoRA fine-tuning is sufficient to produce analyst-quality outputs on mitigation and threat-model reasoning, where strong normative structure exists in the training corpus. Second, performance degrades systematically on tasks requiring operational judgment (e.g., severity estimation and attack-direction classification), highlighting a clear priority for future data augmentation. Third, cosine similarity provides a complementary diagnostic signal; it captures semantic alignment (e.g., 0.832 in the detection signals case) while the LLM judge identifies correctness gaps that similarity alone cannot resolve, enabling more fine-grained model analysis. 

Several scope and evaluation choices contextualize the above findings: The Gemma-4-31B judge introduces known LLM-as-a-judge biases~\cite{zheng2023judging}, which are partially constrained via fixed prompt structure, regex-based parsing with retries, and BGE-based consistency checks, but not fully eliminated. Thus, judge scores should be interpreted as calibrated relative measures rather than ground truth. Inference is intentionally restricted to a minimal metadata block, reflecting the target setting of analyst-grade triage without PCAP access, which partially explains reduced performance on operational judgment tasks. NetsLab flow labels are treated as consistency signals rather than absolute ground truth to avoid propagating labeling uncertainty into training. Finally, evaluation is limited to 313 Fronthaul DDoS replay scenarios on a single 5G/O-RAN testbed, ensuring reproducibility but constraining generalization across interfaces and attack families.

\section{Conclusions}
\label{sec:conclusions}
This paper presented AutoSec-RAN, a specification-grounded LLM framework for automated security validation of AI-native O-RAN. By combining two-stage LoRA fine-tuning with RAG over a curated corpus of O-RAN, 3GPP, and related standards, AutoSec-RAN enables Mistral-7B-Instruct to generate
structured security triage from minimal scenario metadata without requiring raw PCAP access during inference. Evaluated on a physical 5G/O-RAN testbed at Mississippi State University, the framework achieved a mean judge overall score of 0.713 and a blended score of 0.778, performing strongest on mitigation and threat-model reasoning (0.849 and 0.848, respectively) while exposing a controlled training data limitation for attack direction reasoning (0.515). To support reproducibility, the specification index, synthetic question-answer manifests, LoRA adapters, and evaluation harness will be released and made available to the community~\cite{softtue}. 
Beyond the fronthaul DDoS scenarios studied here, the proposed synthetic-pair generation and two-stage LoRA workflow provide a reusable blueprint for specification-grounded LLM adaptation across additional O-RAN interfaces (e.g., F1, E2, and near-RT RIC) and related standards-driven security domains. Future work will validate the judge scores against expert human review, extend AutoSec-RAN to broader O-RAN interfaces, augment RAG with live telemetry summaries, and integrate the framework into the Soft T-UE testing loop~\cite{softue} for continuous security validation and training updates. Together, these extensions will advance AutoSec-RAN to a security analysts with automated triage support for production AI-native RAN operations.

\section*{Acknowledgment}
This work was supported in part by the National Telecommunications and Information Administration (NTIA) under Award No.~28‑60‑IF012 and the National Science Foundation (NSF) Award No.~2555349.
\balance

\bibliographystyle{IEEEtran}
\bibliography{aly.bib}

\end{document}